\documentstyle[12pt,epsf]{article}

\def\hybrid{\topmargin -20pt    \oddsidemargin 0pt
        \headheight 0pt \headsep 0pt
        \textwidth 6.25in       
        \textheight 9.5in       
        \marginparwidth .875in
        \parskip 5pt plus 1pt   \jot = 1.5ex}

\hybrid

\def\baselinestretch{1.2}

\catcode`\@=11

\def\marginnote#1{}
\newcount\hour
\newcount\minute
\newtoks\amorpm
\hour=\time\divide\hour by60
\minute=\time{\multiply\hour by60 \global\advance\minute by-\hour}
\edef\standardtime{{\ifnum\hour<12 \global\amorpm={am}%
        \else\global\amorpm={pm}\advance\hour by-12 \fi
        \ifnum\hour=0 \hour=12 \fi
        \number\hour:\ifnum\minute<10 0\fi\number\minute\the\amorpm}}
\edef\militarytime{\number\hour:\ifnum\minute<10 0\fi\number\minute}

\def\draftlabel#1{{\@bsphack\if@filesw {\let\thepage\relax
   \xdef\@gtempa{\write\@auxout{\string
      \newlabel{#1}{{\@currentlabel}{\thepage}}}}}\@gtempa
   \if@nobreak \ifvmode\nobreak\fi\fi\fi\@esphack}
        \gdef\@eqnlabel{#1}}
\def\@eqnlabel{}
\def\@vacuum{}
\def\draftmarginnote#1{\marginpar{\raggedright\scriptsize\tt#1}}

\def\draft{\oddsidemargin -.5truein
        \def\@oddfoot{\sl preliminary draft \hfil
        \rm\thepage\hfil\sl\today\quad\militarytime}
        \let\@evenfoot\@oddfoot \overfullrule 3pt
        \let\label=\draftlabel
        \let\marginnote=\draftmarginnote
   \def\@eqnnum{(\theequation)\rlap{\kern\marginparsep\tt\@eqnlabel}%
\global\let\@eqnlabel\@vacuum}  }

\def\preprint{\twocolumn\sloppy\flushbottom\parindent 2em
        \leftmargini 2em\leftmarginv .5em\leftmarginvi .5em
        \oddsidemargin -.5in    \evensidemargin -.5in
        \columnsep .4in \footheight 0pt
        \textwidth 10.in        \topmargin  -.4in
        \headheight 12pt \topskip .4in
        \textheight 6.9in \footskip 0pt
        \def\@oddhead{\thepage\hfil\addtocounter{page}{1}\thepage}
        \let\@evenhead\@oddhead \def\@oddfoot{} \def\@evenfoot{} }

\def\numberbysection{\@addtoreset{equation}{section}
        \def\theequation{\thesection.\arabic{equation}}}

\def\underline#1{\relax\ifmmode\@@underline#1\else
        $\@@underline{\hbox{#1}}$\relax\fi}

\def\titlepage{\@restonecolfalse\if@twocolumn\@restonecoltrue\onecolumn
     \else \newpage \fi \thispagestyle{empty}\c@page\z@
        \def\thefootnote{\fnsymbol{footnote}} }

\def\endtitlepage{\if@restonecol\twocolumn \else \newpage \fi
        \def\thefootnote{\arabic{footnote}}
        \setcounter{footnote}{0}}  

\catcode`@=12
\relax

\def\figcap{\section*{Figure Captions\markboth
        {FIGURECAPTIONS}{FIGURECAPTIONS}}\list
        {Figure \arabic{enumi}:\hfill}{\settowidth\labelwidth{Figure
999:}
        \leftmargin\labelwidth
        \advance\leftmargin\labelsep\usecounter{enumi}}}
 \relax
\def\tablecap{\section*{Table Captions\markboth
        {TABLECAPTIONS}{TABLECAPTIONS}}\list
        {Table \arabic{enumi}:\hfill}{\settowidth\labelwidth{Table
999:}
        \leftmargin\labelwidth
        \advance\leftmargin\labelsep\usecounter{enumi}}}
 \relax
\def\reflist{\section*{References\markboth
        {REFLIST}{REFLIST}}\list
        {[\arabic{enumi}]\hfill}{\settowidth\labelwidth{[999]}
        \leftmargin\labelwidth
        \advance\leftmargin\labelsep\usecounter{enumi}}}
 \relax
\makeatletter
\newcounter{pubctr}
\def\publist{\@ifnextchar[{\@publist}{\@@publist}}
\def\@publist[#1]{\list
        {[\arabic{pubctr}]\hfill}{\settowidth\labelwidth{[999]}
        \leftmargin\labelwidth
        \advance\leftmargin\labelsep
        \@nmbrlisttrue\def\@listctr{pubctr}
        \setcounter{pubctr}{#1}\addtocounter{pubctr}{-1}}}
\def\@@publist{\list
        {[\arabic{pubctr}]\hfill}{\settowidth\labelwidth{[999]}
        \leftmargin\labelwidth
        \advance\leftmargin\labelsep
        \@nmbrlisttrue\def\@listctr{pubctr}}}
 \relax
\makeatother
\newskip\humongous \humongous=0pt plus 1000pt minus 1000pt

\newif\ifdtup

\relax

\def\be{\begin{equation}}
\def\ee{\end{equation}}
\def\ba{\begin{eqnarray}}
\def\ea{\end{eqnarray}}

\def\del{\partial}

\def\r{\rho}
\def\a{\alpha}

\def\d{\delta}
\def\D{\Delta}
\def\e{\epsilon}

\def\th{\theta}

\def\m{\mu}
\def\n{\nu}

\def\Om{\Omega}

\def\s{\sigma}

\def\no{\noindent}

\def\qq{\qquad}

\def\IR{\relax{\rm I\kern-.18em R}}

\def \ha {{1\over 2}}

\def \ov {\over}

\def\IR{\relax{\rm I\kern-.18em R}}
\def\inv{^{\raise.15ex\hbox{${\scriptscriptstyle -}$}\kern-.05em 1}}

\def\tL{{\tilde L}}

\begin{document}

\renewcommand{\theequation}{\arabic{equation}}

\newcommand{\beq}{\begin{equation}}
\newcommand{\eeq}[1]{\label{#1}\end{equation}}
\newcommand{\ber}{\begin{eqnarray}}
\newcommand{\eer}[1]{\label{#1}\end{eqnarray}}
\newcommand{\eqn}[1]{(\ref{#1})}
\begin{titlepage}
\begin{center}

\hfill CERN-TH/99-81\\
\hfill hep--th/9903201\\

\vskip .8in

{\large \bf Rotating NS5-brane solution and its exact \\
string theoretical description\footnote{Contribution to the proceedings of the 
{\em 32nd International Symposium Ahrenshoop on the Theory of Elementary
Particles}, Buckow, Germany, 1-5 September 1998 (invited talk).}}

\vskip 0.6in

{\bf Konstadinos Sfetsos}
\vskip 0.1in
{\em Theory Division, CERN\\
     CH-1211 Geneva 23, Switzerland\\
{\tt sfetsos@mail.cern.ch}}\\
\vskip .2in

\end{center}

\vskip .6in

\centerline{\bf Abstract }

\no
We construct the most general solution 
in type-II string theory that represents $N$ coincident 
non-extremal rotating NS5-branes and determine the relevant thermodynamic
quantities. We show that in the field theory limit, 
it has an exact description. 
In particular, it can be obtained by an $O(3,3)$ duality
transformation on the exact string background for the
coset model $SL(2,\IR)_{-N}/U(1) \times SU(2)_N$.
In the extreme supersymmetric limit we recover the multicenter solution, 
with a ring singularity structure, that has been discussed recently.

\vskip 0,2cm

\vskip 4cm
\noindent
CERN-TH/99-81\\
March 1999\\
\end{titlepage}
\vfill
\eject

\def\baselinestretch{1.2}
\baselineskip 16 pt
\noindent

\def\tT{{\tilde T}}
\def\tg{{\tilde g}}
\def\tL{{\tilde L}}


\section{Introduction}

Based on the D-dimensional Kerr solution 
\cite{MP} and its generalization to a family of rotating, electrically 
charged black holes in \cite{CVY1}, a number of solutions
with maximum number of rotational parameters
in 11- and 10-dim supergravities were constructed. Among them in particular,  
the most general solutions representing $N$ coincident rotating M2- or
M5- or D3-branes \cite{CVY2}-\cite{RS}. 
However, an analogous solution representing $N$ coincident
rotating NS5-branes has not been explicitly constructed.
It is the purpose of this note to fill this gap.
It turns out that, because of the absence of R--R fields,
in the near-horizon limit there is a description in terms
of a background corresponding to the exact conformal field theory (CFT) 
$SL(2,\IR)_{-N}/U(1) \times SU(2)_N$. This generalizes previous realizations 
that such exact 
string backgrounds exist in the near-horizon limit 
of $N$ coincident extremal \cite{rey}-\cite{AFK} 
and non-extremal \cite{MS1} NS5-branes, as well 
as for $N$ extremal NS5-branes distributed uniformly along the 
circumference of a ring \cite{KSh}.

\section{Rotating NS5-branes}

The usual NS5-brane solution (extremal or not; see, for instance, \cite{HoSt})
with no angular parameters has a global $SO(4)$ symmetry. 
Introducing angular momentum breaks this symmetry to 
the Cartan subalgebra of $SO(4)$, which is $U(1)\times U(1)$.
Since the latter
is two-dimensional we may obtain a solution with at most two angular 
parameters $l_1$ and $l_2$.
As we shall see, without loss of 
generality, these can be taken to be non-negative.
In order to obtain our solution we have used as a guide 
the general rotating M5-brane
solution \cite{CVY2,CRST}.\footnote{It turns out that 
\eqn{meett}--\eqn{diilla} 
correspond to a dimensional reduction of the M5-brane 
solution we mentioned, along a vanishing circle corresponding to one 
of the angular variables, after we also replace the
mass parameter as $m\to m r$. In particular, this empirical rule can
be used in eq. (2.1) of \cite{CRST} and eq. (14) of \cite{CVY2} for the 
metric and 3-form respectively. The angular variable we mentioned is
denoted by $\psi$ (in both papers) and we dimensionally reduce around
$\psi={\pi/ 2}$.}
The metric of our solution is given by 
\ba
&& ds^2  =  -h dt^2 + dy_1^2 +\dots + dy_5^2
\nonumber \\
&&\phantom{xx}\ 
+ f \left( {dr^2\ov \tilde h} + r^2 ( \D d\th^2 + \sin^2\th \D_1 
d\phi_1^2 + \cos^2\th \D_2 d\phi_2^2)\right) 
\label{meett} \\
&&\phantom{xx}\ 
+ {4 m l_1 l_2 \sin^2\th \cos^2\th\ov r^2 \D} d\phi_1 d\phi_2
-{4 m \cosh\a \ov r^2 \D}\ dt 
(l_1 \sin^2\th d\phi_1 + l_2 \cos^2\th d\phi_2) \ ,
\nonumber
\ea
the components of the antisymmetric tensor by
\ba
B_{\phi_1\phi_2}&  =&  -2 m \cosh \a \sinh \a \left(1+{l_1^2\ov r^2}\right)
{\cos^2\th\ov \D}\ ,
\nonumber \\
B_{t\phi_1} & = & 2 m l_2 \sinh\a {\sin^2\th \ov r^2 \D}\ ,
\label{antiis}\\
B_{t\phi_2} & = & 2 m l_1 \sinh\a {\cos^2\th \ov r^2 \D}\ ,
\nonumber
\ea
and the dilaton by 
\ba
e^{2\phi} = g_s^2 f \ ,
\label{diilla}
\ea
where $g_s $ is the string coupling at infinity\footnote{The general rotating
$D5$-brane solution in type-IIB supergravity is trivially 
obtained by an S-duality 
transformation on \eqn{meett}--\eqn{diilla} and will not be presented here.}
and the various functions are defined as
\ba
f& =&  1+ {2m \sinh^2\a \ov r^2 \D} \ ,\qq h\ =\ 1 - {2 m\ov r^2 \D}\ ,
\nonumber \\
\tilde h & =&  {1\ov \D} \left( 1+ {l_1^2\ov r^2} + {l_2^2\ov r^2}
+ {l_1^2 l_2^2\ov r^4} - {2 m\ov r^2} \right) \ ,
\nonumber \\
\D & =&  1 +  {l_1^2\ov r^2}\cos^2\th  + {l_2^2\ov r^2}\sin^2\th\ ,
\label{deeff}\\
\D_1 &  =& 
 1 +  {l_1^2\ov r^2}  + {2 m l_1^2 \sin^2\th\ov r^4\D f}\ ,
\nonumber \\
\D_2 &  =& 
 1 +  {l_2^2\ov r^2}  + {2 m l_2^2 \cos^2\th\ov r^4\D f}\ .
\nonumber
\ea
The ADM mass, the angular momentum and the angular velocities associated with
motion in $\phi_1$ and $\phi_2$, as well as the Bekenstein--Hawking entropy 
and temperature are given by\footnote{The angular velocities 
$\Om_i$, $i=1,2$ in \eqn{MJS} below, are determined by demanding that 
the three-vector (with components in the
$t,\phi_1$ and $\phi_2$ directions) $\eta^a=(1,\Om_1,\Om_2)$ be null
at the horizon, i.e. $\eta^2|_{r_H}=0$. The temperature is determined using
the general formula $T_H^2 =-{1\ov 16 \pi^2} \lim_{r\to r_H}
{\nabla_\m \eta^2 \nabla^\m \eta^2 \ov \eta^2}$.}
\ba
M_{ADM}& =&  {\Om_3 V_5 \ov 16 \pi G_N} 2 m (2 \cosh^2 \a + 1)\ ,\quad
\Om_3 = 2 \pi^2\ ,
\nonumber \\
J_i & =&  {\Om_3 V_5 \ov 4 \pi G_N}  m l_i  \cosh\a\ ,\qq i=1,2\ ,
\nonumber\\
\Om_i& =& {l_i\ov (r_H^2 + l_i^2)\cosh\a}\ , \qq i=1,2\ ,
\label{MJS}\\
S & =&  {\Om_3 V_5 \ov 4 G_N} 2  m r_H  \cosh\a\ ,
\nonumber\\
T_H &= &{r_H^4-l_1^2l_2^2\ov 4\pi m r_H^3 \cosh\a}\ ,
\nonumber
\ea
where $r_H$ is the position of the outer horizon given by 
\be
r_H^2 = \ha \left(2m -l_1^2 -l_2^2 + 
\sqrt{(2m -l_1^2 -l_2^2)^2- 4 l_1^2 l_2^2} \right)\ .
\label{hoorr}
\ee
There is also an inner horizon given by the above formula with a minus sign
in front of the square root. 
Notice also that in order to have a horizon, i.e. $r_H^2\geq 0$,
the inequality $l_1+l_2\leq \sqrt {2 m}$ should be satisfied.
The parameter $\a$ is related to the mass and charge of the NS5-brane by
\be
\sinh^2 \a = \sqrt{\left({\a' N\ov 2 m}\right)^2 + {1\ov 4} } - \ha\ .
\label{smasn}
\ee
Finally we note that the thermodynamic quantities in \eqn{MJS} obey
the first law of black-hole thermodynamics
\be
dM_{ADM} = T_H dS + \Om_1 dJ_1 + \Om_2 dJ_2\ .
\label{thee}
\ee
This is easily checked by treating $M_{ADM},S,J_1,J_2$ as functions
of the variables $m,l_1,l_2$ using \eqn{MJS}--\eqn{smasn}.

\section{The extremal limit}

The extremal limit of the above solution is obtained by letting $m\to 0$.
Then, after changing variables from 
$(r,\th,\phi_1,\phi_2)$ to $(x_1,x_2,x_3,x_4)$ as 
\be
\pmatrix{ x_1 \cr x_2} =  \sqrt{r^2+l_1^2}\ \sin\th \pmatrix{\cos\phi_1\cr
\sin\phi_1}\  ,\qq \pmatrix{x_3\cr x_4}=  \sqrt{r^2+l_2^2}\ \cos\th 
\pmatrix{\cos\phi_2\cr \sin\phi_2}\ ,
\label{x1234}
\ee
we find the following background
\ba
&&ds^2=-dt^2 + dy_1^2 +\dots + dy_5^2 + 
H dx_i dx_i\ ,\qq i=1,2,3,4\ ,
\nonumber \\
&& H_{ijk}= \e_{ijkl} \del_l H\ ,
\label{axins}\\
&& e^{2\Phi}= H\ ,
\nonumber
\ea
where $H$ is 
given by 
\be 
H = 1 + { \a' N \ov \sqrt{(l_1^2-l_2^2 + x_1^2+ x_2^2+ x_3^2 + x_4^2)^2
- 4 (l_1^2-l_2^2) (x_1^2 +x_2^2)} } \ .
\label{hhh} 
\ee 
It can easily be checked that $H$ is a (multicenter) harmonic 
function in the 
4-dim Euclidean space spanned by the $x_i$'s.
The metric in \eqn{hhh} has singularities at 
\ba
&& x_3 = x_4 = 0 \ ,\qq x_1^2+x_2^2 = l_1^2-l_2^2\ ,\quad {\rm if}\ \
l_1>l_2\ ,
\nonumber \\
&& x_1 = x_2 = 0 \ ,\qq x_3^2+x_4^2 = l_2^2-l_1^2\ ,\quad {\rm if}\ \ 
l_1<l_2\ .
\label{siing}
\ea
Hence, the singularity structure is that of a ring with radius $\sqrt{|l_1^2
-l_2^2|}$. In fact, \eqn{axins} with \eqn{hhh} 
corresponds to a 
continuous uniform distribution of 
NS5-branes along the circumference of a ring \cite{KSh}.\footnote{My 
interest in finding the rotating NS5-brane solution \eqn{meett}--\eqn{diilla} 
was sparked by the (correct) remark of E. Kiritsis that 
the BPS solution \eqn{axins}, could be unstable at finite temperature,
since the gravitational attraction will no longer be balanced by just the
R--R repulsion. In our solution \eqn{meett}--\eqn{diilla}
spin forces provide the necessary extra balance.}
In the field-theory limit, discussed in \cite{KSh}, the 1 in the harmonic 
function in \eqn{hhh} is effectively removed. Then, it becomes an exact string 
background as it is connected by a T-duality
transformation to the coset model CFT $SL(2,\IR)_{-N}/U(1)\times
SU(2)_N/U(1)$ \cite{KSh}.

The background \eqn{axins} is an axionic instanton and as such it preserves
half of the supersymmetries of flat space. 
From a gauge theory view point, 
it corresponds to a Higgs phase of a 6-dim SYM theory $SU(N)$
broken to $U(1)^N$ since the centers where the branes are put correspond
to non-zero expectation values for the scalars.
In our case the vacuum moduli space has a $Z_N\times U(1)$ symmetry,
which, in the continuous limit we are
discussing here, becomes a $U(1)\times U(1)$ symmetry.
This degeneracy is however lifted once we turn on the temperature,
and the corresponding supergravity solution can describe excitations around
these points of the moduli space.

\section{Field-theory limit and exact description }

A natural question arises,
namely what the field-theory limit of the non-supersymmetric background 
\eqn{meett}--\eqn{diilla} is and, moreover if it also has an exact CFT 
interpretation as well.
Consider the limit $g_s\to 0$ and 
$m\to 0$ in such a way that the ratio $m^{1/2}/g_s $ is held fixed.
In this limit the Yang--Mills coupling constant 
$g_{\rm YM}\sim \a'$ remains finite. 
It is convenient to define rescaled quantities as
\ba 
{2 m\ov g_s^2 } =  \m \a'\ , \qq
r = (2 m)^{1/2} \r\ ,\qq l_i = (2 m)^{1/2} a_i \ ,\quad i=1,2 \ ,
\label{defgk}
\ea
and then take the linit $m\to 0$ in (\ref{meett})--(\ref{diilla}). We find
for the metric\footnote{In the following we use the rescaled variables 
$t\to \sqrt{\a' N} t$ and $y_i\to \sqrt{\a' N} y_i$, $i=1,\dots , 5$, and 
omit $\a'$ since it drops out of the $\s$-model as well as the supergravity 
action.}
\ba
{1\ov N} ds^2 & =& -\left(1-{1\ov \D_0} \right) dt^2 + dy_1^2+ \dots +dy_5^2 
+  {d\r^2 \ov \r^2 + a_1^2 a_2^2/\r^2 + a_1^2+a_2^2-1} 
\nonumber \\
&& +\ d\th^2 +  {1\ov \D_0} \left( ( \r^2 + a_1^2) \sin^2\th d\phi_1^2 
+ ( \r^2 + a_2^2) \cos^2\th d\phi_2^2 \right)
\label{fgha}\\
&&-\ {2\ov \D_0}\ dt\ (a_1 \sin^2\th d\phi_1 + a_2 \cos^2\th d\phi_2)\ ,
\nonumber
\ea 
for the antisymmetric tensor two-form 
\be 
{1\ov N} B =  2 {1\ov \D_0}\left(- (\r^2 + a_1^2) \cos^2\th  d\phi_1 \wedge
d\phi_2 + a_2 \sin^2\th  dt \wedge d\phi_1 + a_1 \cos^2\th  
dt \wedge d\phi_2 \right)\ ,
\label{fgha1}
\ee
and for the dilaton 
\be
e^{2\Phi} = {N\ov \m \D_0 } \ . 
\label{lakla}
\ee
The function $\D_0$ entering the previous expressions is defined as
\be
\D_0= \r^2 + a_1^2 \cos^2\th + a_2^2 \sin^2\th \ .
\label{dsfg}
\ee
Note that string-theory corrections to the supergravity result 
are organized in powers of $1/N$.
Hence, by choosing $N\gg1$ we suppress these 
perturbative corrections. On the other 
hand, string-loop corrections are suppressed by choosing $N\ll \m$. These are 
the same conditions as were 
found in \cite{MS1} for the case of zero angular momenta.
As a final remark we note that it is very likely that the background 
\eqn{fgha}--\eqn{lakla} can also be obtained by gauging directly a 
2-dim subgroup, isomorphic to $U(1)\times U(1)$, of the WZW model 
for $SL(2,\IR)\times SU(2)$. In that case we may compute the 
${1\ov N}$-corrections to the background \eqn{fgha}--\eqn{lakla} using 
techniques developed in \cite{BaSfexa}

\subsection{The $O(3,3)$ duality transformation}

First, consider the case of vanishing angular parameters $a_1$ and $a_2$.
Then, the background \eqn{fgha}-\eqn{lakla} becomes the one 
corresponding to the $SL(2,R)_{-N}/SO(1,1)
\times SU(2)_N$ exact CFT, as it was shown in \cite{MS1}.
It turns out that by performing an $O(3,3)$ transformation to the latter 
background we can obtain the more general one given by \eqn{fgha}-\eqn{lakla}.
Let us first pass to the Euclidean regime by letting $t\to -i \tau$ 
and $a_1\to i a_1$. In order to find out the specific $O(3,3)$ matrix,
we first expand the $\s$-model action with metric and 
antisymmetric tensor given by \eqn{fgha} and \eqn{fgha1} for small values
of $a_1, a_2$. Then, the infinitesimal change in the $\s$-model 
Lagrangian density is  
\ba
\d {\cal L} & =&  - {a_1-a_2\ov\cosh^2 r}\ ( \sin^2\th 
\del_+ \tau \del_- \phi_1 -\cos^2\th \del_+ \tau \del_- \phi_2) 
\nonumber \\
&& - {a_1+a_2 \ov \cosh^2 r}\ (\sin^2\th \del_+ \phi_1\del_- \tau +
 \cos^2\th \del_+ \phi_2 \del_- \tau) + {\cal O}(a^2)\ ,
\label{def1d}
\ea
where we have changed variables as $\r=\cosh r$ so that $G_{rr}=1$ to 
zeroth order in $a_1,a_2$.
In the space of the three variables $X^\m=(\tau,\phi_1,\phi_2)$ a general 
$O(3,3)$ transformation acts as (see for instance \cite{GPR})
\be
\tilde E = (a E+ b)(c E + d)\inv\ ,
\label{oddd}
\ee
where the group element $G=\pmatrix{a & b\cr c & d}\in O(3,3)$ preserves the 
bilinear form $J=\pmatrix{0 & I\cr I & 0}$, i.e. $G^T J G= J$.
The matrix $E_{\m\n}= G_{\m\n}+ B_{\m\n}$ is read off from the 
Euclidean version of the background \eqn{fgha}, \eqn{fgha1} (after setting 
$a_1=a_2=0$):
\be
E= \pmatrix{ \tanh^2 r & 0 & 0 \cr 0 & \sin^2\th & -\cos^2\th \cr 
0 & \cos^2\th & \cos^2\th}\ .
\label{eegl}
\ee
An infinitesimal version of the transformation \eqn{oddd} 
is obtained by expanding the $O(3,3)$ group element around the identity 
element using 
$a=I + A$, $b= B$, $c= C$ and $d=I -  A^T$, where $B$ and $C$ 
are antisymmetric matrices. Then, the infinitesimal change (first order 
in the generators $A$, $B$ and $C$) of the $\s$-model Lagrangian density is 
\be
\d {\cal L}= (A E + E A^T + B - ECE)_{\m\n} \del_+ X^\m \del_- X^\n .
\label{djl}
\ee
Comparing \eqn{def1d} and \eqn{djl} we determine 
\be
A=\pmatrix{0 & -a_1 & -a_2\cr a_1 & 0 & 0\cr 0 & 0 &0 }\ , 
\quad
B= \pmatrix{0 & a_2 & 0\cr -a_2 & 0 & 0\cr 0 & 0 &0 }\ , 
\quad
C=\pmatrix{0 & a_2 & a_1\cr -a_2 & 0 & 0\cr -a_1 & 0 & 0 }\ .
\label{abcw}
\ee
Exponentiating, we find that the necessary $O(3,3)$ group element in \eqn{oddd}
is
\ba
a & = & \pmatrix{ \s_1 \s_2 & -b_1\s_1\s_2 & -b_2 \s_1 \s_2 \cr
b_1\s_1 \s_2& \s_1\s_2& -b_1 b_2 \s_1 \s_2\cr
0 & 0 & 1}\ ,\quad
b\ =\ \pmatrix{ b_1 b_2 \s_1 \s_2 & b_2\s_1\s_2 & 0 \cr
- b_2\s_1 \s_2& b_1 b_2 \s_1\s_2& 0\cr
0 & 0 & 0}\ ,
\nonumber \\
c & =& \pmatrix{ b_1 b_2 \s_1 \s_2  &b_2 \s_1 \s_2 &  b_1\s_1 \s_2 \cr
-b_2 \s_1 \s_2 & b_1 b_2 \s_1 \s_2 & 1- \s_1 \s_2 \cr
- b_1 \s_1 \s_2 & 1- \s_1 \s_2 & b_1 b_2 \s_1 \s_2}\ ,\quad
d \ = \ \pmatrix{ \s_1 \s_2 & -b_1\s_1\s_2 & 0 \cr
b_1\s_1 \s_2& \s_1\s_2& 0\cr
b_2 \s_1 \s_2 & -b_1 b_2 \s_1\s_2 & 1}\ ,
\label{exll} 
\ea
where 
\ba
&&\s_i^2\equiv {\r_+^2 -a_i^2\ov \r_+^2-\r_-^2}\ ,\quad b_i^2\equiv 
{a_i^2 -\r_-^2\ov \r_+^2-a_i^2}\ ,\qq i =1,2\ ,
\nonumber\\
&& \r_\pm^2 \equiv 
\ha\left(a_1^2+a_2^2+1 \pm \sqrt{(a_1^2+a_2^2+1)^2 - 4 a_1^2 a_2^2}\right)\ .
\label{exll1}
\ea
Indeed, we may easily check that applying \eqn{oddd}, with \eqn{eegl} and 
\eqn{exll}, we obtain a matrix $\tilde E$; after we change variables as
$\r^2 = (\r_+^2-\r_-^2)\cosh^2 r + \r_-^2$, this $\tilde E$
corresponds to the Euclidean
version of the background \eqn{fgha} and \eqn{fgha1}. The dilaton \eqn{lakla}
is found by demanding that the measure factor $e^{-2\Phi} \sqrt{\det G}$
be invariant under the $O(3,3)$ transformation.

\bigskip\bigskip
\noindent
{\large \bf Acknowledgements}

\smallskip
\noindent
I would like to thank the organizers for the invitation to present this and
related work.

\vfill\eject

\end{document}